\documentclass[preprint,trackchanges]{aastex631}
\usepackage [latin1]{inputenc}
\hypersetup{linkcolor=blue,citecolor=blue,filecolor=cyan,urlcolor=blue}
\shorttitle{Sequential Magnetic Reconnections in a Fishbone-like Structure}
\shortauthors{Yang et al.}

\begin{document}

\title{Sequential Magnetic Reconnections in a Fishbone-like Structure Leading to Recurrent Brightenings}

\correspondingauthor{Bo Yang}
\email{boyang@ynao.ac.cn}

\author{Bo Yang}
\affil{ Yunnan Observatories, Chinese Academy of Sciences, 396 Yangfangwang, Guandu District, Kunming, 650216, People's Republic of China}
\affil{ Yunnan Key Laboratory of Solar Physics and Space Science, Kunming 650011, People's Republic of China}

\author{Jiayan Yang}
\affil{ Yunnan Observatories, Chinese Academy of Sciences, 396 Yangfangwang, Guandu District, Kunming, 650216, People's Republic of China}
\affil{ Yunnan Key Laboratory of Solar Physics and Space Science, Kunming 650011, People's Republic of China}

\author{Junchao Hong}
\affil{ Yunnan Observatories, Chinese Academy of Sciences, 396 Yangfangwang, Guandu District, Kunming, 650216, People's Republic of China}
\affil{ Yunnan Key Laboratory of Solar Physics and Space Science, Kunming 650011, People's Republic of China}

\author{Yi Bi}
\affil{ Yunnan Observatories, Chinese Academy of Sciences, 396 Yangfangwang, Guandu District, Kunming, 650216, People's Republic of China}
\affil{ Yunnan Key Laboratory of Solar Physics and Space Science, Kunming 650011, People's Republic of China}



\begin{abstract}
The fine-scale release of magnetic free energy in the solar atmosphere is a fundamental open question in solar physics. 
Multi-wavelength observations at high spatiotemporal resolution now offer a direct window into this process.
Using data from NVST, \emph{SDO}, \emph{IRIS}, and \emph{Hinode}, 
we reveal the energy release process in a fishbone-like magnetic structure within active region 12297.
The fishbone-like structure consists of a spine along a narrow, elongated positive-polarity field, with herringbone branches rooted in negative-polarity sunspots. 
Persistent photospheric magnetic flux emergence and shearing motions are observed beneath the fishbone-like structure, 
which may play a key role in maintaining its topology and producing the recurrent brightenings.
During brightenings, compact bright features propagate sequentially from west to east along the spine, accompanied by bidirectional flows along the branches and plasma blobs ejected toward the distant positive sunspot. 
These propagating features could be interpreted as signatures of sequential magnetic reconnection events occurring in chronological order at nodes along the spine, predominantly in the chromosphere and transition region.
Our observations may provide evidence that recurrent brightenings could arise from repeated sequential reconnection events organized by a coherent magnetic structure,
deepening our understanding of how such recurrent brightenings are generated and how magnetic free energy is dissipated at fine scales in solar active regions.
\end{abstract}


\section{Introduction}  \label{sec:intro}
The accumulation and release of magnetic energy in the solar atmosphere serve as the fundamental physical mechanism driving magnetic activities across a wide range of spatiotemporal scales, 
from nanoflares and microflares to X-class major flares. Understanding these processes has long been a central topic in solar physics, 
as it is essential for explaining coronal heating \citep{parker83,parker88,klim06} and the triggering of solar eruptive activities \citep{chen11,shibata11,cheng17}. 
However, the underlying physical mechanisms and fine-scale processes of energy release remain poorly understood. 
Local brightenings, which occur frequently at various scales in the solar atmosphere, represent key signatures of this energy release. 
These brightenings arise when magnetic free energy stored in specific field structures is rapidly released through mechanisms such as magnetic reconnection \citep{parker57,lin2000,li16,xue16,ding22,yan22}.
The released energy is subsequently converted into plasma kinetic energy, thermal energy, and high-energy particle kinetic energy, leading to a significant enhancement of local radiation.
Therefore, tracking and analyzing the detailed evolution of these brightenings using high-resolution observations 
may provide crucial insights into the fine-scale processes of magnetic free energy release.

Clarifying the morphology and evolution of the fine magnetic structures that store magnetic free energy is essential.
It is widely accepted that both magnetic flux emergence \citep{oka08,yan17} and magnetic reconnection driven by photospheric motions can form magnetic flux ropes of various scales \citep{van89,yang16,cheng17,yang21,shen24}, 
and that the eruption of these structures can produce flares of different classes except nanoflares. To explain small-scale energy release,
\citet{parker83,parker88} proposed that random photospheric convective motions can continuously shuffle and stretch the footpoints of coronal magnetic fields, 
forcing them to intertwine and wrap around one another. This process inevitably leads to the formation of a braided magnetic structure,
within which numerous small-scale current sheets form due to magnetic field misalignment,
where magnetic free energy is impulsively released via small-scale, self-organized cascading reconnection events, producing storms of nanoflares. 
Parker's theory predicts that such braided structures along with their evolutionary processes should be abundant in the solar corona and, in principle, observable. 
However, observational confirmation has proven challenging. On one hand, these braided structures are intrinsically small in scale and become quickly untied by magnetic reconnection in their early stages; 
the accompanying heat spreads rapidly along magnetic strands, rendering the structures difficult to resolve  \citep{sch07,pontin17,peter22}. 
On the other hand, \citet{asch19} pointed out that braided topologies cannot account for the observed stability or low twist number of coronal loops in a force-free corona, 
suggesting that Parker-type nanoflaring may only occur in non-force-free environments, such as the chromosphere and transition region.

To date, only a few observations have successfully captured the braided structures and partially resolved the evolutionary processes that produce nanoflares.
Using extreme ultraviolet (EUV) and X-ray observations of an active region, \citet{pare10} studied the relaxation and multithermal evolution of tangled loops in a postflare arcade. 
Employing high spatiotemporal coronal EUV images from the High-resolution Coronal Imager \citep{kob14}, 
\citet{cir13} identified signatures of spatially resolved magnetic braids and associated coronal heating in two low-lying loops within an active region. 
Subsequent studies building on this work further investigated the fine-scale magnetic structure and evolution of these braids \citep{tha14,tiw14,bi23}. 
One of the braids was found to correspond to a low-lying twisted flux rope above a penumbral filament region \citep{tha14}.
Notably,  the other example of small braided loops, shown in Figure 3 of \citet{cir13}, exhibits a distinctive fishbone-like structure.
With new high-resolution observations from the Extreme Ultraviolet Imager (EUI) aboard \emph{Solar Orbiter} \citep{muller20}, 
\citet{chit22} provided recent evidence for the untangling of small-scale coronal braids in the core of active regions, a process associated with either gentle or impulsive heating of the plasma in coronal loops. 
More recently, \citet{chen25} presented direct imaging evidence of magnetic reconnection during the untangling of braided magnetic structures above a sunspot. 
In a separate study focusing on an activated prominence,  \citet{chen20} observed very fast, bursty minijets occurring roughly perpendicular to the prominence magnetic field. 
These tiny jet phenomena were termed ``nanojets" by \citet{antolin21}, who interpreted them as a consequence of magnetic reconnection and the subsequent relaxation of braided coronal loops.
In addition, \citet{huang18} reported that magnetic braids are not only associated with coronal heating but can also directly drive eruptions.
These observations have successfully identified braided structures and captured glimpses of isolated reconnection events within them.
However, the predicted dynamic cascade of numerous interconnected, small-scale reconnection events, often referred to as magnetic avalanches,
that drive nanoflare storms has not yet been directly observed.

Introduced by \citet{bak87,bak88}, the concept of self-organized criticality (SOC) describes a nonlinear dynamical system driven toward a critical state independent of external parameters, 
in which a local instability occurs when a local parameter exceeds a threshold, and the resulting energy release triggers a transport process that may cause neighboring sites to also surpass the threshold, 
leading to a cascade of instabilities and thus producing intermittent avalanches with power-law frequency distributions.
\citet{lu91} proposed that the solar coronal magnetic field resides in such an SOC state and transformed Parker's coronal heating model \citep{parker83,parker88} into an avalanche model for flares of all sizes.
In their framework, random photospheric motions stress the coronal magnetic field, and local magnetic energy increases first trigger nanoflares, 
which subsequently induce nearby nanoflares, producing a cascading energy release that grows from small to large flares.
In a complementary study, \citet{berger09} examined reconnection in braided magnetic fields and showed that such systems self-organize to a critical state with power-law distributions in coherence lengths and energy release, supporting the broader SOC framework for coronal heating. Notably, the model favors smaller, more frequent energy releases near loop ends.
For decades, magnetic avalanche models rooted in SOC have been widely used to explain the power-law distributions of hard X-ray (HXR) peak fluxes observed in solar flares, 
providing a robust statistical link between flare occurrence rates and avalanche dynamics \citep[for reviews, see][]{char01,asch16}. 
Nevertheless, detailed studies of the fine-scale energy release processes underlying magnetic avalanches have remained scarce.
The current development of high-resolution numerical simulations and high spatiotemporal resolution observations now makes it possible to probe the fine-scale physical processes of magnetic avalanches.
Using a three-dimensional magnetohydrodynamic (MHD) simulation, \citet{hood16} demonstrated how the instability and reconnection of one magnetic thread 
can trigger neighboring threads to become unstable, thereby initiating an MHD avalanche.
Critically, high-resolution observations from Solar Orbiter have revealed that a solar flare is driven by an avalanche-like process, 
in which initially weak but rapid reconnection events (on timescales of seconds) progressively trigger more prominent activity, ultimately leading to an explosive flare  \citep{chit26}. 
This discovery unveils the central engine of the flare and highlights the essential role of magnetic avalanches in enabling impulsive, bursty energy release in the solar corona. 
Consequently, these advances in high-resolution simulations and observations are opening  a new era for understanding the fine-scale dynamics of magnetic energy release, 
shifting the focus from statistical descriptions to direct physical explanations of individual avalanche processes.

In this Letter, we present the detailed evolution of recurrent brightenings within a fishbone-like structure in an active region,
utilizing high spatiotemporal resolution data from the New Vacuum Solar Telescope \citep[NVST;][]{liu14}, the \emph{Solar Dynamics Observatory} \citep[\emph{SDO};][]{pes12},
the \emph{Interface Region Imaging Spectrograph} \citep[\emph{IRIS};][]{de14}, and the \emph{Hinode} \citep{kos07}.
High-resolution observations reveal bright features moving along the spine of the fishbone-like structure during these brightenings. 
We interpret these moving features as tracing a sequence of magnetic reconnection events occurring in chronological order at the nodes along the spine,
with the entire process taking place predominantly within the chromosphere and transition region.
This study may offer observational evidence  that recurrent brightenings  could arise from repeated sequential reconnection events organized by a coherent magnetic structure,
significantly advancing our understanding of the fine-scale processes governing magnetic free energy release.

\section{Observations}
The recurrent brightenings in the fishbone-like structure occurred in NOAA active region (AR) 12297 on 2015 March 11-12. 
They were  well observed by the Atmospheric Imaging Assembly \citep[AIA;][]{lem12} and the Helioseismic and Magnetic Imager \citep[HMI;][]{sch12} on board the \emph{SDO}.
The AIA instrument takes the full-disk images of the Sun in seven extreme-ultraviolet (EUV) and two ultraviolet (UV) wavelengths with a pixel size of 0$\arcsec$.6 at a high cadence of up to 12s.
Here, the Level 1.5 images centered at 304 \AA \ (\ion{He}{2}, 0.05 MK), 171 \AA\ (Fe {\sc ix}, 0.6 MK ), 193 \AA\ (Fe {\sc xii}, 1.3 MK and Fe {\sc xxiv}, 20 MK), 131 \AA\ (Fe {\sc viii}, 0.6 MK and Fe {\sc xxi}, 10 MK),
and  94 \AA\ (\ion{Fe}{18}, 7 MK) were adopted to study the event. The HMI instrument provides full-disk continuum intensity images and the vector magnetic field data with a pixel size of 0.$\arcsec$5.
The time cadence of the HMI are 45 s (for the continuum intensity images) and 720 s (for vector magnetic field data).
The AIA data used here cover from 2015 March 11 21:00 UT to 2015 March 12 06:25 UT, and the HMI data  from 2015 March 11 00:00 UT to 2015 March 12 10:00 UT. 

On 2015 March 12, AR  12297 was also observed by the NVST in $H_{\alpha}$ 6562.8 \AA\ and TiO during  05:34-09:53 UT.
The pixel sizes are approximately 0.$\arcsec$165 for $H_{\alpha}$  images and 0.$\arcsec$052 for TiO images, with temporal cadences of 12 s and 30 s, respectively. 
Raw data were reconstructed into high-resolution images through dark current subtraction, flat-field correction, and speckle masking  \citep{xiang16}. 
Unfortunately, the \emph{SDO} observations suffered a data gap from 06:25 UT to 08:30 UT.
The AR 12297 on 2015 March 11-12 was tracked by \emph{IRIS} slit jaw imager (SJI) in 1330 and 1400 \AA\ . 
Two time intervals of data were used to analyze the detailed evolution of recurrent brightenings in the transition region: the first from 2015-03-11 22:47:58 to 2015-03-12 03:47:15 UT, 
with a spatial resolution of 0.$\arcsec$332 pixel$^{-1}$ and a temporal resolution of 20 s; 
and the second from 2015-03-12 04:56:58 to 05:15:04 UT, with a spatial resolution of 0.$\arcsec$166 pixel$^{-1}$ and a temporal resolution of 68 s.
The same AR and the recurrent brightenings were also observed by the Solar Optical Telescope (SOT) on board \emph{Hinode} in the near-photospheric G-band (4305 \AA)
and in the chromospheric  \ion{Ca}{2} H (3968.5 \AA) channel with a pixel size of 0.$\arcsec$108. 
Data were acquired during two intervals (2015-03-11 19:03-2015-03-12 05:53 UT; 2015-03-12 06:20-10:15 UT).
The G-band has a fixed cadence of 10 minutes, whereas the \ion{Ca}{2} H  cadence is 68 s for the first interval and 60 s for the second.
In addition, an soft X-ray (SXR) image from the X-ray Telescope (XRT) on board the \emph{Hinode} was used to show the appearance of a brightening.
To compensate for the solar rotation, all of the above data were then differentially rotated to the  reference time of 23:30 UT on 2015 March 11.
Finally, all NVST, \emph{IRIS}, and \emph{Hinode} images were mapped to heliographic coordinates and co-aligned with SDO observations using an automatic mapping technique \citep{ji19}.

\section{Results}
\subsection{The Fishbone-like Structure and Associated Recurrent Brightenings}
Figure 1({\it a}) illustrates the general appearance of the AR 12297. From 2015 March 11 21:00 UT to 2015 March 12 09:00 UT, 
a series of recurrent brightenings occurred at the southeastern edge of the AR (see Figure 1({\it a}), ({\it c}), and ({\it g})).
As shown in Figure 1({\it g}), each peak of the curve corresponds to a brightening process.
In the core region where the brightenings occur, a narrow strip of positive polarity, labeled ``p", 
is sandwiched between two negative-polarity sunspots, ``N1" and ``N2''. N2 and p are a newly emerged dipole between the $\delta$ sunspots "P" and N1. 
High-resolution NVST TiO observation (Figure 1(b)) shows that the penumbral fibrils of N1 and N2  in this region are highly sheared and are both connected to p.
Of particular note is that, in the upper atmosphere, specifically the chromosphere and transition region,
high-resolution observations reveal a fishbone-like structure aligned along the location of p (see Figure 1({\it d}), ({\it e}), and ({\it f})). 
This structure, characterized by a long, narrow bright band that forms the spine and a series of herringbone-shaped branches extending outward from both sides of the spine to connect to N1 and N2, 
bears a strong resemblance to the braided structure described and illustrated in Figure 3 of \citet{cir13}.
Additionally, nodal structures are located at the intersections between the branches and the spine (Figure 1({\it f})).
These nodes are likely the core sites of magnetic reconnection and might often be accompanied by energy release and bidirectional plasma flows.
The observed brightenings all first appeared in the region where N1, N2, and p came into contact with each other, corresponding to the location of the fishbone-like structure. 
They then expanded and extended along the spine of the fishbone-like structure, steadily intensifying to a peak within minutes before slowly fading away.
These recurrent brightenings are likely driven by repeated magnetic reconnection among the magnetic fields within the fishbone-like structure.

To determine the recurrence time intervals and occurrence frequency of the brightening events, we analyzed the SDO/AIA 193\,\AA\ light curve of the
target region. The raw light curve was smoothed with a Savitzky--Golay filter \citep{sav64} (window length $= 51$ points, polynomial order $= 3$) to suppress high-frequency noise. 
The detection threshold was set at $T = B + 3\sigma_{\mathrm{rms}}$, where the background level $B$ is the 5th percentile of the smoothed flux, 
and $\sigma_{\mathrm{rms}}$ is the rms noise of the quiet intervals. Brightening events were identified using a peak-finding algorithm with the criteria: (i)~peak height $\geq T$, 
(ii)~minimum separation $\geq 50$ data points ($\sim 10.5$\,min), and (iii)~prominence $\geq 80$\,DN\,s$^{-1}$. Closely spaced peaks (e.g., events 9 and 14) 
were merged into a single event, yielding a final catalogue of 18 distinct brightening episodes. 
The duration of each event was measured as the Full Width at Half Maximum (FWHM) on the smoothed curve.
The waiting time, also known as the recurrence interval, is defined as the time difference between the peak moments of two adjacent brightening events.

From this analysis, we obtain a mean brightening duration (FWHM) of $\sim 9.0$\,min. The recurrence time intervals range from $13.2$\,min to $59.2$\,min, 
with a mean of $\sim 29.7$\,min and a median of $26.2$\,min (Figure 2({\it b})), corresponding to an event occurrence rate of $\sim 1.91$\,hr$^{-1}$.
The light curve (Figure 2({\it a})) shows that brightening events are distributed throughout the observing window, with a cluster near 00:00--01:30\,UT. 
The waiting-time histogram (Figure 2({\it c})) reveals a broad distribution spanning nearly a factor of five, lacking strict periodicity. 
This behavior is consistent with the SOC paradigm \citep{bak87,berger09}, in which coronal energy accumulation and release proceed via scale-free avalanches, 
yielding a broadened waiting-time distribution rather than a single periodic modulation.

\subsection{Ejection of Plasma Blobs During Brightening Events}
By scrutinizing the observations from the \emph{SDO}/AIA, plasma blobs are frequently observed to be ejected from the source region of the brightenings. 
Figure 3 shows four selected brightening events during which plasma blob ejections were observed.
The first such event occurred on 2015 March 11. At about 22:32 UT, brightening initially appeared in the mixed-polarity region formed by the negative-polarity sunspots N1 and N2 
and the narrow positive-polarity strip p between them (Figures 3({\it a})). The brightening region then gradually expanded, accompanied by the ejection of plasma blobs (Figures 3({\it b}) and ({\it c})). 
The arrows in Figures 3({\it b}) and ({\it c}) track the motion of this blob. The second (panels ({\it d} \sbond {\it f})), third (panels ({\it g} \sbond {\it i})), 
and fourth (panels ({\it j} \sbond {\it l})) brightening events presented here all occurred on 2015 March 12.
Their origins, morphologies, and evolutionary processes are very similar to those of the first event. 
Specifically, the brightening first appeared in the mixed-polarity region formed by N1, N2, and P.
As it expanded or evolved, plasma blobs were ejected from it and moved toward the positive-polarity sunspot region P.
These brightening processes, accompanied by plasma ejections, are likely related to magnetic reconnection within the fishbone-like structure. 
It is conceivable that the magnetic field lines connecting P to N2 and those connecting P to N1 intertwine near the negative-polarity footpoints N1 and N2, 
thereby forming part of the fishbone-like structure. These field lines might undergo reconnection, producing the observed brightenings and driving plasma blobs toward the positive-polarity footpoint P. 

\subsection{The Fine Process of Brightening Evolution}
On 2015 March 12, AR 12297 was also observed by NVST, \emph{Hinode}, and \emph{IRIS}, all of which captured fragments of the brightening events of our interest. 
The high-resolution chromospheric and transition region data provided by these instruments offer an unprecedented opportunity 
to  trace the key details of brightening formation and evolution within the fishbone-like magnetic structure.
As shown in Figure 4,  two selected brightening events for which the detailed evolutionary process can be clearly traced.
Similar to the brightening events described above, both brightening events here first appeared at the mixed-polarity region formed by N1, N2, and p (Figure 4({\it a1}), ({\it b1}), and ({\it c1})). 
The first  brightening appeared at about 00:08 UT (Figure 4({\it a1}), ({\it b1}), and ({\it c1})) and the second one at about 00:27 UT (Figure 4({\it a4}), ({\it b4}), ({\it c4}), and ({\it d4})).
They then expanded from west to east along the location of p, with their intensity increasing simultaneously, and the entire process lasted for several minutes (Figure 4({\it b1} \sbond {\it b3}) and Figure 4(({\it c1} \sbond {\it c3})).
The AIA 131 observations reveal that the core of each brightening was located exactly at the negative-polarity footpoints of the coronal loops connecting P\sbond N1 and P\sbond N2, 
and as the brightenings developed, the emission from these loop structures also intensified (Figure 4({\it a1} \sbond {\it a6})).
Furthermore, during the development of the brightenings, bright ribbons were also observed to propagate from west to east along the negative-polarity magnetic fields N1 and N2, 
with the ribbon along N1 being more pronounced (Figure 4({\it c1} \sbond {\it c6})). 
It is evident from the \emph{IRIS} 1330 \AA\ images that magnetic loops along the spine of the fishbone-like structure are clearly connected to these ribbons.
The generation mechanism of these ribbons is similar to that of flare ribbons, and they may result from sequential magnetic reconnection along the spine of the fishbone-like structure from west to east.

The most prominent feature during the development of the brightenings is a compact bright structure moving from west to east along the spine of the fishbone-like structure 
(Figure 4({\it d1} \sbond {\it d3}) and Figure 4(({\it d4} \sbond {\it d6})). With this motion, both the brightenings and the footpoint ribbons gradually expanded and moved from west to east.
 High-resolution NVST chromospheric observations reveal further details of how this bright structure moves along the fishbone-like structure, 
leading to the generation and expansion of brightenings as well as the formation and evolution of footpoint bright ribbons.
Figure 5 presents the detailed evolution of a brightening process observed by NVST.
At around 06:36 UT on 2015 March 12, a compact bright structure began to appear at the western segment of the fishbone-like structure's spine (indicated by arrows in Figures 5({\it a}), ({\it e}), and ({\it i})). 
At this time, faint footpoint brightenings appeared in the regions of the negative-polarity magnetic fields N1 and N2.
As the compact bright structure moved from west to east along the spine of the fishbone-like structure, 
the footpoint bright ribbon along N1 first developed from west to east and then turned northward, forming a hook-like flare ribbon (Figures 5({\it b} \sbond {\it d}), and ({\it f} \sbond {\it g})).
In contrast, the bright ribbon on the N2 side remained relatively compact and showed no significant apparent motion. 
Additionally, a bright ribbon also appeared at the distant positive-polarity region P, which developed from west to east and then turned southward,
forming another hook-like flare ribbon (Figures 5({\it c} \sbond {\it d}) and ({\it g})).
High-resolution NVST observations further reveal that, during the west-to-east motion of the compact bright structure, 
on the one hand, material moved from the bright structure toward both sides, tracing the magnetic loops connecting the bright structure to the flare ribbons on either side (Figures 5({\it b} \sbond {\it d}) and  ({\it f} \sbond {\it g})).
On the other hand, a material flow directed toward the positive-polarity magnetic field P along the fishbone-like structure was also observed (indicated by cyan arrows in Figures 5({\it b}) and ({\it f})).
To track the dynamic evolution of the compact bright feature, a spacetime plot was constructed along slice ``AB'' in Figure 5({\it c}) using NVST $H_{\alpha}$ line center images, 
and the result was provided in Figure 5({\it m}). We find that the compact bright feature moved along the spine of the fishbone-like structure from west to east with a mean velocity of about 16.9 km s$^{-1}$.
These speeds are consistent with the expected Alfv\'{e}n speed in the chromosphere and transition region \citep{chae23}, 
supporting the interpretation that the moving feature traces successive reconnection sites rather than a bulk plasma flow.
During the NVST observing interval, a data gap occurred in SDO, whereas \emph{Hinode}/XRT captured several rare coronal frames. 
The \emph{Hinode}/XRT image shows that the morphology of this brightening event is nearly identical to that described previously: 
the source region exhibits strong emission, accompanied by slightly diffuse coronal loops that extend from the source region to the positive-polarity sunspot P (Figure 5({\it h})).

The westward-to-eastward motion of the compact bright structure along the spine of the fishbone-like structure, and the consequent formation and evolution of brightenings within this structure, 
closely resemble the process of unzipping a zipper. Notably, this bright structure moving along the spine was captured only in the chromospheric observations of NVST and \emph{Hinode}/SOT, 
and it likely represents a locally enhanced radiative feature occurring in the chromosphere. How is it generated, and what physical process does its motion reveal? 
Based on our observations, we infer that this compact bright structure may result from a sequence of magnetic reconnection events occurring from west to east, in chronological order, at the nodes of the fishbone-like structure. 
Thus, the bright structure likely traces the sites of magnetic reconnection, which primarily take place in the chromosphere.
According to our earlier inference, the magnetic field lines connecting P\sbond N1 and P\sbond N2 may intertwine near the negative-polarity footpoints, forming part of the fishbone-like structure. 
This inference is consistent with the conclusions of \citet{asch19} and \citet{berger09}, who suggest that intertwined magnetic structures are more prevalent in the chromosphere or transition region, 
corresponding to the footpoint regions of coronal loops.
Ultimately, the series of magnetic reconnections occurring within the fishbone-like structure not only generate the moving compact bright structure and the flare ribbons distributed along N1, N2, and P, 
but also naturally give rise to bidirectional material flows toward both sides of the fishbone-like structure, as well as material flows and plasma blobs directed along the spine toward the positive-polarity magnetic field P.
The question of whether these sequential reconnections are causally linked, i.e., whether reconnection at one node triggers reconnection at neighboring nodes, is discussed in Section~4.
Thanks to the high-resolution observations from NVST, \emph{Hinode}, and \emph{IRIS}, 
our observations reveal the detailed, fine-scale process by which sequential reconnection along the fishbone-like magnetic structure leads to energy release and the subsequent formation of brightenings.

\subsection{Magnetic Reconnection at a Single Node on the Spine of the Fishbone-like Structure}
The sequential reconnections along the spine of the fishbone-like structure drive the propagating bright features.
Reconnection at a single node serves as the fundamental building block of the entire sequential reconnection process.
Here, we examine individual reconnection events at isolated nodes. Figure 6 presents two representative cases of magnetic reconnection occurring at a single node.
The first event began at around 01:51:18 UT on 2015 March 12. A compact bright point first appeared on the spine of the fishbone-like structure (Figure 6({\it a})), 
exactly at the node where its herringbone branches converge. Subsequently, bidirectional material flows emerged from this bright point (Figures 6({\it b} \sbond {\it c})), 
tracing the herringbone branches of the structure more clearly (Figures 6({\it d})).
The second event occurred on the same day and exhibited similar evolutionary behavior. At around 03:21:57 UT, 
a bright point appeared on the spine of the fishbone-like structure (Figure 6({\it e})), followed by bidirectional material flows from the bright point (Figures 6({\it f} \sbond {\it h})).
AIA EUV observations of these events show that the emission from the corresponding coronal regions was significantly enhanced during their evolution (Figures 6({\it i} \sbond {\it l})).
These observations suggest that magnetic fields forming the herringbone branches of the fishbone-like structure may undergo reconnection at the nodes, 
thereby producing bidirectional material flows, heating the local regions, and leading to the observed emission enhancements.

\subsection{Photospheric Magnetic Field Evolution Underlying the Fishbone-like Structure}
Because the recurrent brightenings occurred in the fishbone-like structure, which corresponds to the opposite-polarity magnetic flux region beneath it, 
the underlying photospheric magnetic field evolution in this region should contain the key clues to revealing the formation and maintenance of the fishbone-like structure, 
as well as the generation and evolution of the recurrent brightenings. Through checking the HMI data, 
we found that significant flux emergence commenced between the  $\delta$ sunspots P and N1 before the appearance of the fishbone-like structure.
This is shown by the HMI vertical images in Figures 7({\it a} \sbond {\it f}).
This flux emergence process persisted throughout our entire observing interval.
As is common behavior for emerging magnetic flux, the positive and negative poles of the emerging magnetic flux moved away from each other (panels ({\it a} \sbond {\it d})).
The positive pole p moved northwest and then approached the negative sunspot N1. 
The negative pole moved southeast and then merged with the negative sunspot N. 
During this process, the negative polarity field ``n" on the western side of the newly emerged dipole also moved toward the negative pole of the newly emerged dipole and N, 
together forming a larger negative sunspot N2 (panels ({\it c} \sbond {\it e}) and ({\it g} \sbond {\it i})). Eventually, p evolved into a narrow, elongated structure sandwiched between N1 and N2.
Although p, together with N1 and N2, formed a mixed-polarity region, flux emergence continued throughout our entire observing interval, making it difficult for us to determine whether magnetic cancellation occurred between them.

It is notable that the transverse fields, which are represented by the short red and blue arrows with the arrow length proportional to the relative field strength and the alignment parallel to the field direction, 
were changed and enhanced significantly on the side where p and N1 contacted (panels ({\it c}) and ({\it f})).
These arrows were distributed along p and oriented toward N1 (see inserted image in panel ({\it f})). 
Furthermore, high-resolution photospheric observations revealed highly sheared penumbral fibrils that also formed along p during this process and extended into sunspot N1 (panels ({\it h} \sbond {\it i})), 
following the same pattern as the transverse fields. These observations suggest that magnetic cancellation might occurred between p and N1 during their evolution, thereby establishing new magnetic connectivity.
Applying the inductive DAVE4VM \citep{sch08} method to the 12 minute cadence HMI vector magnetic field data, 
we calculated the photospheric velocity field, which integrated over 2 hr between 16:12 UT to 18:12 UT on 2015 March 11 and superimposed on an HMI vertical image (panel ({\it d})).
Remarkably, persistent southeastward photospheric flows characterize the negative sunspots N1 and N. 
Meanwhile, p moved northwestward throughout its emergence, although this motion was not resolved by photospheric flow tracking.
Together, they displayed overall strong shearing motion over the core region of the brightenings.
As flux emergence, photospheric shearing motions, and likely magnetic cancellation persisted throughout the formation and maintenance of the fishbone-like structure and the generation of the recurrent brightenings,
 we therefore speculate that these persistent processes may sustain the stressed, braided topology of the fishbone-like structure and continuously supply the free energy required for repeated sequential reconnections.
 
\section{Conclusion and Discussion}
In this study, we presented high-resolution observations of sequential magnetic reconnections driving recurrent brightenings in a fishbone-like structure.
The fishbone-like structure, revealed by chromospheric and transition region imaging, consists of a spine aligned with a narrow, 
elongated positive-polarity field and herringbone-shaped branches rooted in negative-polarity sunspots. Persistent photospheric flux emergence, shearing motions, 
and probable magnetic cancellation beneath the structure likely sustain its complex topology and supply free energy for the recurrent brightenings.
The key finding is that, during the brightenings, compact bright features propagate sequentially along the spine, 
likely tracing a series of magnetic reconnection events occurring in chronological order at the nodes in the chromosphere or transition region.
This process is accompanied by bidirectional flows along the branches and plasma blob ejections along the spine directed toward the distant positive-polarity sunspot.
These observations demonstrate that the brightenings are not isolated events but rather part of a coordinated, sequential energy release process organized by the fishbone-like magnetic structure. 
This study may significantly advance our understanding of how persistent photospheric driving and repeated sequential small-scale reconnection generate recurrent brightenings, 
highlighting that high-resolution observations are essential for probing the fundamental building blocks of magnetic energy release.

As described in Section 3.1, the fishbone-like structure is the central physical entity in our observations.
This structure might be an active site of sequential magnetic reconnection. 
Its topology strongly resembles the braided magnetic configurations predicted by \citet{parker88} and recently imaged in the context of coronal heating \citep{cir13,chit22}. 
However, unlike previous studies that focused on coronal braids, our fishbone-like structure is predominantly observed in the chromosphere and transition region,
consistent with the suggestion by \citet{asch19} that such intertwined topologies are more likely to survive in non-force-free environments. 
Thus, the fishbone-like structure could serve as a key observational target for understanding small-scale energy release in the lower solar atmosphere.

Persistent photospheric magnetic flux emergence, shearing motions, and likely magnetic cancellation are observed beneath the fishbone-like structure throughout the entire observing interval. 
These processes may drive the continuous supply of magnetic free energy and maintain the complex topology of the structure. 
Specifically, the newly emerged dipole (N2 and p) interacts with the pre-existing sunspots N1 and P, leading to highly sheared penumbral fibrils and enhanced transverse fields. 
These conditions might be ideal for the formation of a braided or fishbone-like configuration. 
The recurrent brightenings originate exactly from this mixed-polarity region and propagate along the spine. 
Therefore, we propose that the fishbone-like structure may  act as a preferential site for repetitive magnetic reconnection, driven by persistent photospheric magnetic flux emergence and shearing motions. 
This is distinct from previous studies of one-time braided loop untangling \citep[e.g.,][]{cir13,chit22,bi23} and instead represents a recurring energy release engine in active regions.

The most striking feature of our observations is the propagation of a compact bright structure from west to east along the spine of the fishbone-like structure, 
accompanied by the sequential development of footpoint ribbons and bidirectional flows. 
This propagation resembles the ``zipping" of a zipper and is naturally interpreted as a sequence of magnetic reconnection events occurring in chronological order at the nodes where branches meet the spine. 
The motion of the bright feature might track the chronological progression of reconnection sites along the spine. 
This interpretation could be  further supported by the observation of bidirectional flows from individual nodes (Figure 6) and the ejection of plasma blobs along the spine toward the positive-polarity sunspot P (Figure 2). 
It is noteworthy that the moving compact bright feature along the spine of the fishbone-like structure 
is clearly detected only in the chromospheric (NVST $H_{\alpha}$, \emph{Hinode}/SOT \ion{Ca}{2} H ) and transition region (IRIS 1330 \AA, 1400 \AA) observations, 
but is absent or significantly blurred in coronal EUV images. This may suggest that the sequential reconnection events, as well as the resulting propagating bright front, 
are primarily confined to the chromosphere and transition region.

The entire evolution of a brightening event, from the initial appearance of a compact bright feature at the west of the spine to its eastward propagation, 
the sequential brightening of ribbons along N1, N2, and P, and the accompanying material flows, appear to resemble the magnetic avalanche process introduced by  \citet{lu91}.
A key question is whether the observed sequential reconnections along the spine are causally connected, specifically, whether  
reconnection at one node redistributes the local magnetic stress sufficiently to destabilize neighboring regions, thereby triggering the next event in a propagating cascade.
In our observation, the continuous propagation of the bright feature along the spine at a speed of $\sim$16.9 km s$^{-1}$ (as measured in Figure 6(m)),
which is commensurate with the local Alfv\'{e}n speed, together with the lack of appreciable temporal gaps,  might be suggestive of a cascading energy release.
However, we emphasize that we have not directly observed reconnection at one node triggering reconnection at neighboring nodes.
Our inference is based on the observed propagation of bright features and the chronological development of ribbons and flows, not on direct measurement of causal triggering between individual nodes.
Whether this process genuinely constitutes an avalanche-like chain reaction, or whether the sequential brightenings are independently driven by a common external trigger 
such as progressive photospheric shearing and flux emergence, remains an open question.
Our observations could directly link the recurrent brightenings of the active region to repeated sequential reconnection activity along the spine of the fishbone-like structure.

We also note that alternative interpretations should be considered.
The plasma blob ejections observed along the spine may be associated with plasmoid-mediated reconnection \citep[e.g.,][]{li16,yang18,yan22}, 
in which multiple plasmoids form within an extended current sheet and are ejected sequentially.
It is possible that plasmoid-mediated reconnection occurred locally at individual reconnection sites during our event.
However, the overall observational characteristics, including the sequential brightenings at distinct nodes along the spine and the bidirectional flows perpendicular to the spine along the herringbone branches,
cannot be fully accounted for by plasmoid-mediated reconnection alone, which primarily describes dynamics within a single current sheet rather than an organized sequential reconnection chain along an extended structure.
Another possibility is slipping reconnection, in which the reconnection site appears to move along the spine as field lines sequentially reconnect at progressively displaced positions within a quasi-separatrix layer.
This could also produce a propagating bright feature without causal cascading, as discussed below.

Slipping reconnection, a three-dimensional reconnection process occurring in quasi-separatrix layers (QSLs) \citep{priest95,demoulin96,titov02,aul06}, offers another possible interpretation for some of our observations.
In this process, field lines ``slip'' through the diffusion region, producing an apparent motion of the reconnection site along the QSL \citep{janvier13,dudik14,li15}.
The west-to-east propagation of the compact bright feature along the spine and the sequential brightening of footpoint ribbons along N1, N2, 
and P could be interpreted as the apparent motion of the reconnection site and the tracing of the QSL footprint on the photosphere, respectively.
However, several observational features favor the interpretation of sequential magnetic reconnection at discrete nodes over a purely slipping reconnection scenario.
Most notably, the bidirectional flows observed perpendicular to the spine along the herringbone branches (Figure 6) are difficult to reconcile with a single, continuously slipping reconnection front, 
as they imply localized, distinct reconnection sites where material is expelled in opposite directions along the branches.
Furthermore, the ejection of plasma blobs along the spine toward the distant positive sunspot P (Figure 3), while possibly associated with local plasmoid-mediated reconnection at individual nodes, 
more strongly supports the picture of sequential magnetic reconnections occurring at distinct sites along the spine rather than a uniformly sliding reconnection front.
We therefore suggest that while slipping reconnection may account for the apparent motion of the bright feature and the ribbon development, 
our observations, particularly the perpendicular bidirectional flows and localized plasma blob ejections, are more consistent with a series of reconnection events occurring at distinct nodes along the spine.
We caution, however, that the observed fishbone-like structures and the associated activities are located in the chromosphere and transition region. 
We cannot rule out the possibility that the apparent motion of the flare ribbons results from slipping reconnection of coronal magnetic fields in QSLs during the brightening.

Recurrent EUV brightenings have been widely reported in the solar atmosphere, often attributed to repeated magnetic reconnection driven by flux emergence or cancellation \citep{jiang07,chif08,gug18,yang23},
persistent reconnection at null points in fan-spine topologies \citep{cheng23}, or continuous bald patch reconnection \citep{pol17}.
However, most previous studies focused on isolated brightenings or simple loop-loop interactions. 
In contrast, our work reveals that recurrent brightenings can be organized by a coherent, extended magnetic structure, the fishbone-like topology, 
and that individual brightening events  could connect through a sequential reconnection chain along the spine.
This provides a new paradigm: recurrent brightenings in active regions may arise not only from repeated external driving 
but also from sequential reconnection organized by a pre-existing network of braided field lines.

Despite the unprecedented detail of our observations, several limitations remain.
Most importantly,  we cannot determine from imaging data alone whether the observed sequential reconnections are causally triggered, 
i.e., whether reconnection at one node directly triggers reconnection at neighboring nodes, or whether they are independently driven by a common external process.
While we observe bidirectional flows from individual nodes (Figure 6) and the propagation of a bright feature along the spine (Figure 5),
the triggering mechanism remains unresolved.
This is partly due to the limited cadence and spatial resolution of the instruments, and the fact that the core reconnection sites are small and evolve on timescales of seconds.
Future high-resolution observations with instruments such as \emph{Solar Orbiter}/EUI (high-cadence mode) \citep{muller20} and  the Daniel K. Inouye Solar Telescope (DKIST) \citep{rast21}
will be essential to resolve the nodal triggering process and to determine whether the sequential reconnections are causally connected, as in the magnetic avalanche scenario, or driven by external processes.

\begin{acknowledgments}
The authors sincerely thank the anonymous referee for detailed comments and useful suggestions for improving this manuscript.
The authors thank the scientific/engineering team of \emph{SDO}, \emph{IRIS}, \emph{Hinode}, and  NVST for providing the excellent data.
This work is supported by the National Key R$\&$D Program of China (2024YFA1612000);
 the Natural Science Foundation of China (12273108, 12273106, and 12203097); the Yunnan Science Foundation of China  (202601AT070305 and  202501AT070025); 
the Yunnan Province XingDian Talent Support Program; the ``Yunnan Revitalization Talent Support Program" Innovation Team Project (202405AS350012);
the CAS ``Light of West China" Program; and the Yunnan Key Laboratory of Solar Physics and Space Science (202205AG070009).
\end{acknowledgments}

\newpage
\begin{figure}
\epsscale{1.0}
\plotone{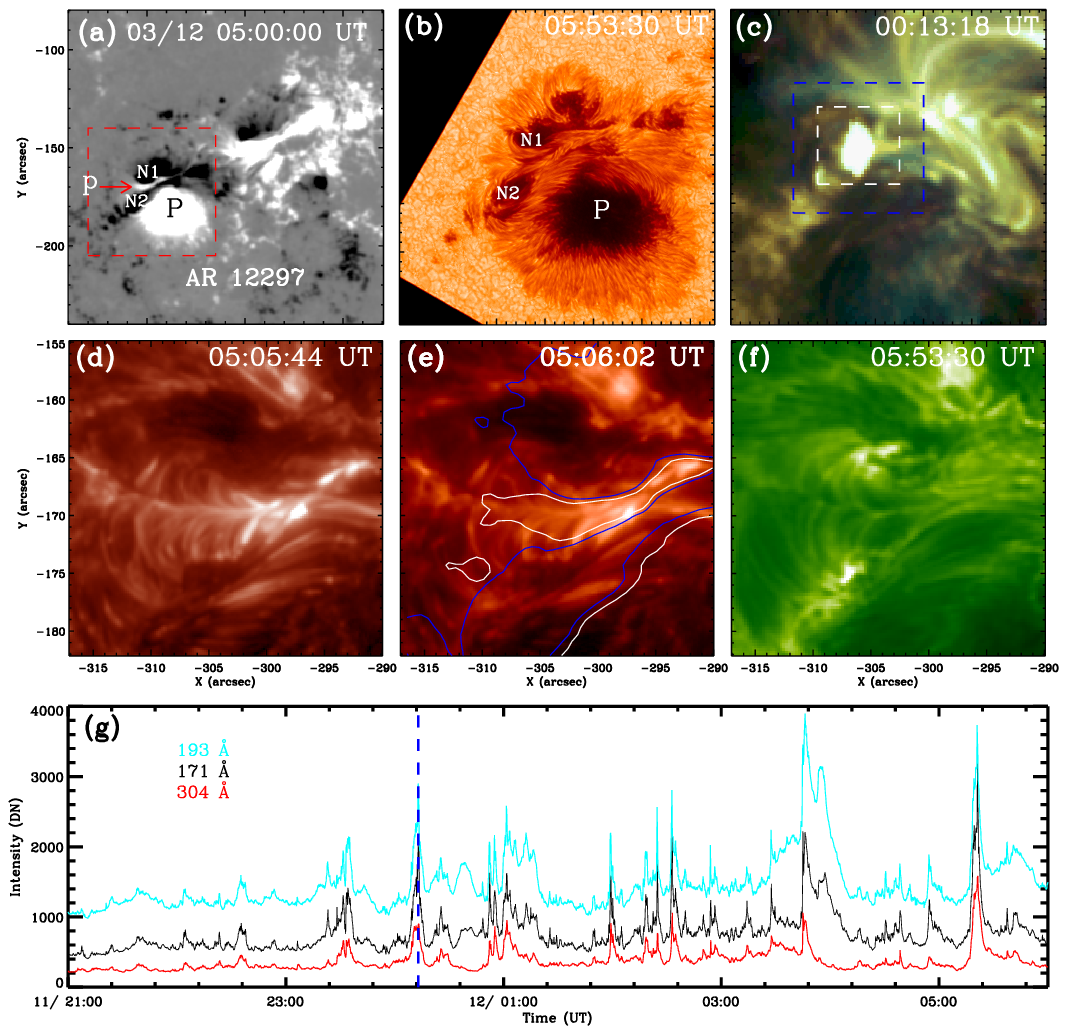}
\caption{Observations of the fishbone-like structure in AR 12297 on 2015 March 12. ({\it a}) Vertical component of the  \emph{SDO}/HMI vector magnetic field at 05:00:00 UT, showing an overview of AR 12297.  
({\it b})  NVST TiO image taken at 05:53:30 UT.  Labels ``N1'' and ``N2'' register the sunspots with negative polarity, 
while ``P'' denotes the sunspot with positive polarity. Label ``p'' indicates a narrow strip of positive polarity sandwiched between N1 and N2, located directly beneath the fishbone-like structure.  
({\it c}) Composite of the AIA 171 \AA\ (red), 193 \AA\  (green), and 131\AA\ (blue)  image shows a typical ongoing brightening event. ({\it d} \sbond {\it f}) \emph{IRIS}/SJI 1330 \AA\ , 1400 \AA\
, and NVST  $H_{\alpha}$  line center images displaying the general appearance of the fishbone-like structure. 
Iso-Gauss contours of  $\pm 200$ G are superposed by white and blue curves in panel ({\it e}).
The red dashed box outlines the FOV of panels ({\it b} \sbond {\it c}), and the blue one outlines the FOV of panels ({\it d} \sbond {\it f}).
({\it g}) AIA 304 \AA\ (red), 171 \AA\ (black), and 193 \AA\ (cyan) light curves extracted from the white dashed rectangle in panel ({\it c}) . 
The vertical dashed line marks the time of the image shown in panel ({\it c}).
}
\end{figure}

\begin{figure}
\epsscale{1.1}
\plotone{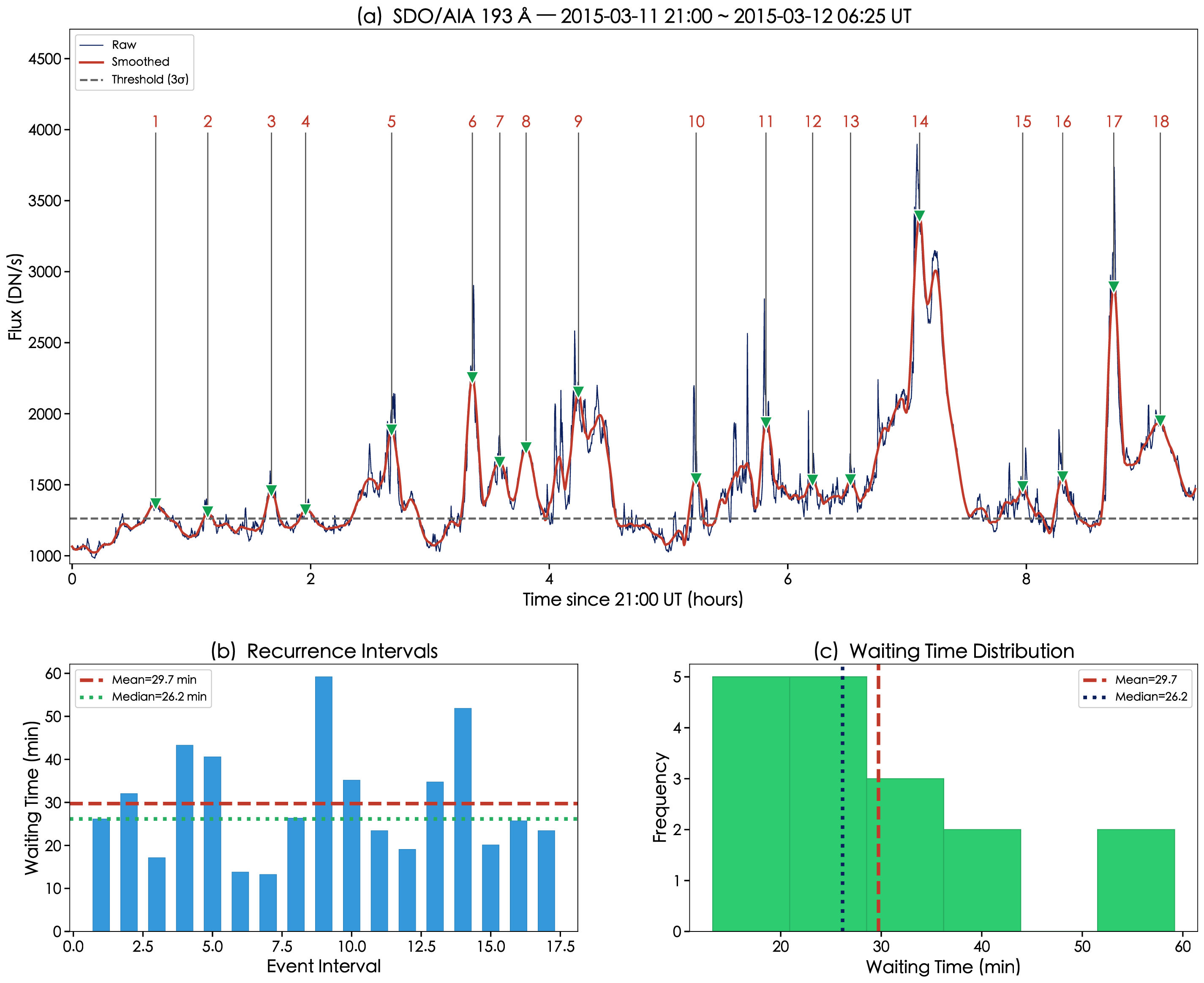}
\caption{ \emph{SDO}/AIA 193\,\AA\ light-curve analysis. ({\it a}) Raw (dark blue) and Savitzky--Golay smoothed (red) ligh curve from 21:00 UT on 2015 March 11 to 06:25 UT on 2015 March 12.
The gray dashed line marks the $3\sigma$ detection threshold. Green inverted triangles indicate the 18 detected brightening events, labeled 1--18 with gray leader lines.
 ({\it b}) Waiting times between consecutive events. Dashed and dotted lines denote the mean and median, respectively.
 ({\it c}) Histogram of the waiting-time distribution with mean (dashed) and median (dotted) overlaid.}
\end{figure}

\begin{figure}
\epsscale{1.2}
\plotone{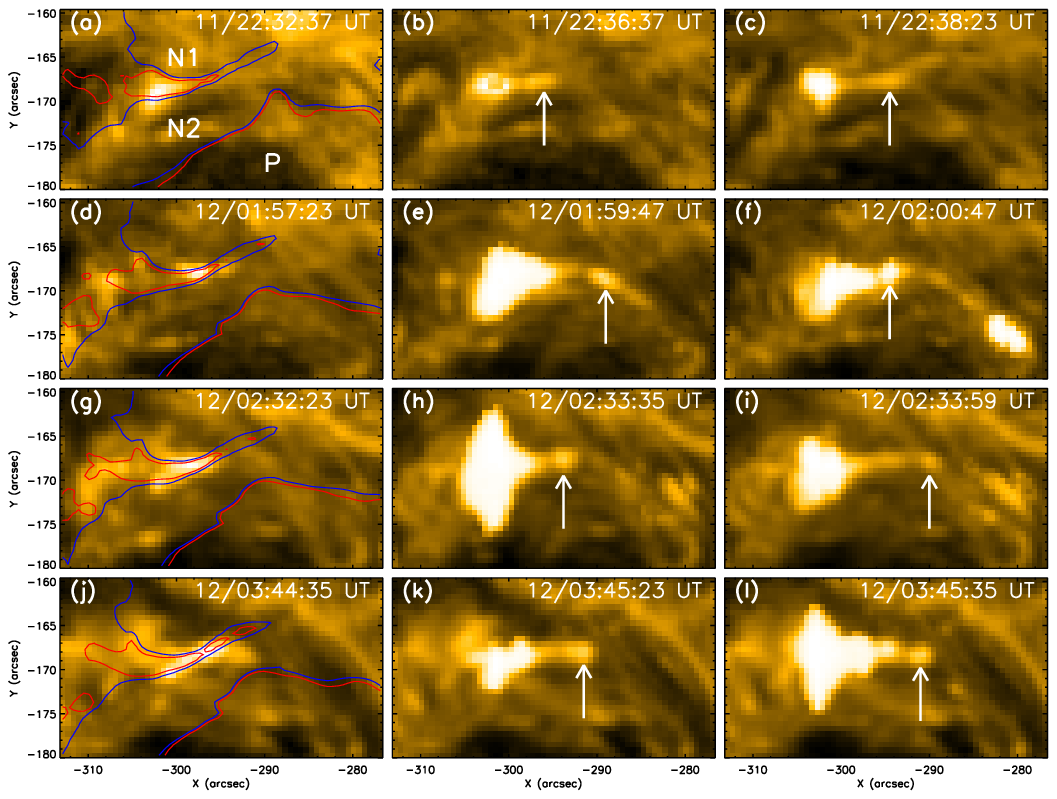}
\caption{({\it a} \sbond {\it l}) Sequence of  AIA 171 \AA\ images displaying the ejection of plasma blobs from the brightenings, 
which themselves originate from the fishbone-like structure and are ejected toward the distant positive sunspot P.
Simultaneous HMI magnetograms  are overplotted on  panels ({\it a}), ({\it d}), ({\it g}), and ({\it j}) as red/blue contours for positive/negative polarity, 
with contour levels of 50 G and -200 G, respectively. The white arrows point to the ejected plasma blobs.
An animation of this figure is available. In the movie, events 1\sbond4 correspond to panels (({\it a} \sbond {\it c}), ({\it d} \sbond {\it f}), ({\it g} \sbond {\it i}), and ({\it j} \sbond {\it l}), respectively.
The animation has a 2 s  cadence.}
\end{figure}

\begin{figure}
\epsscale{1.1}
\plotone{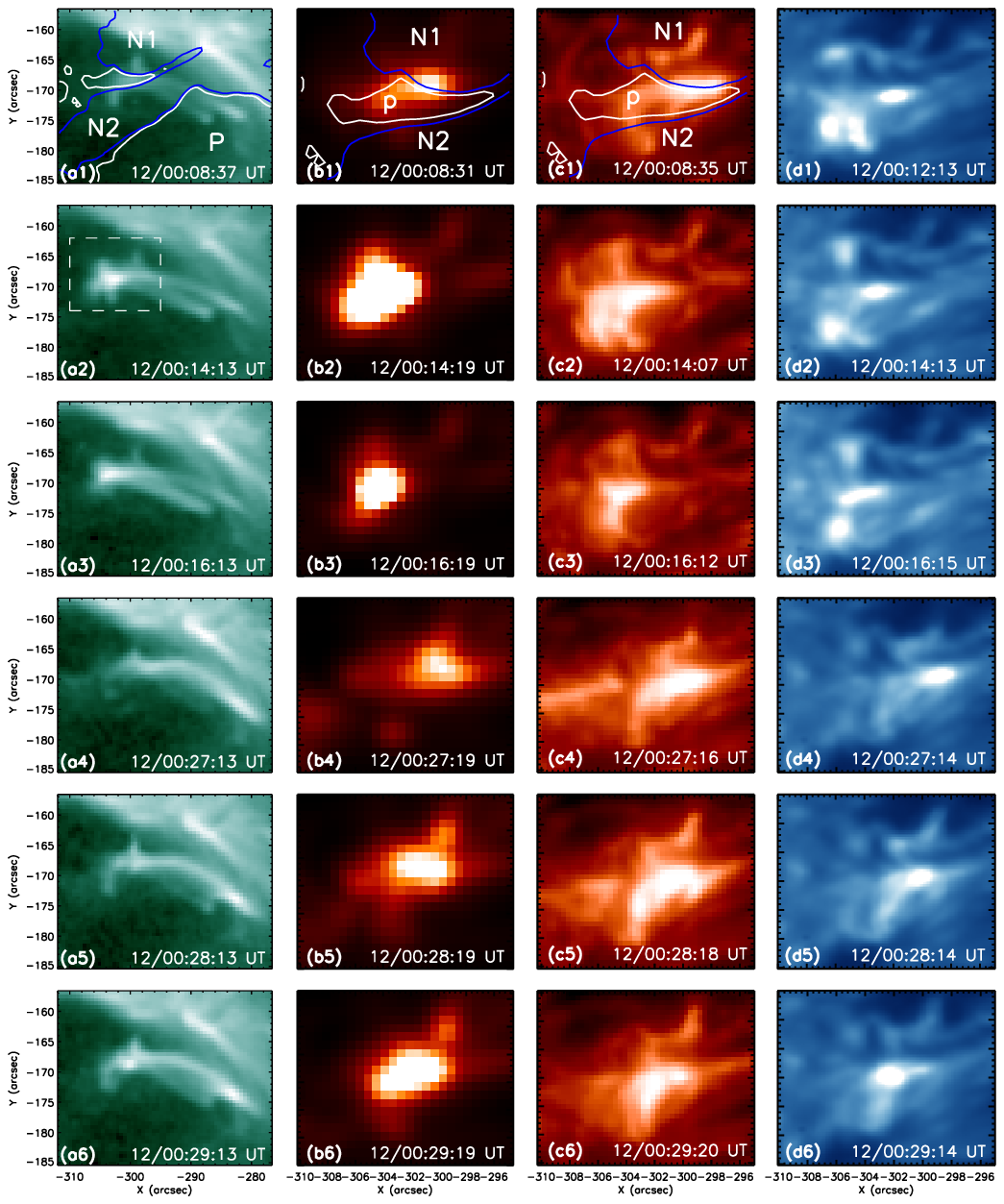}
\caption{AIA 94 \AA\ (panels ({\it a1} \sbond {\it a6})), 304 \AA\ (panels ({\it b1} \sbond {\it b6})), \emph{IRIS} 1330 \AA\ (panels ({\it c1} \sbond {\it c6})), 
and  \emph{Hinode}/SOT  \ion{Ca}{2} H (panels ({\it d1} \sbond {\it d6})) images showing the detailed evolution of two brightening events occurring in the fishbone-like structure. 
The dashed box outlines the FOV of panels ({\it b1} \sbond {\it d6}). Iso-Gauss contours of  100 G and -200G are superposed by white and blue curves in panels ({\it a1}), ({\it b1}), and ({\it c1}).}
\end{figure}

\begin{figure}
\epsscale{1.1}
\plotone{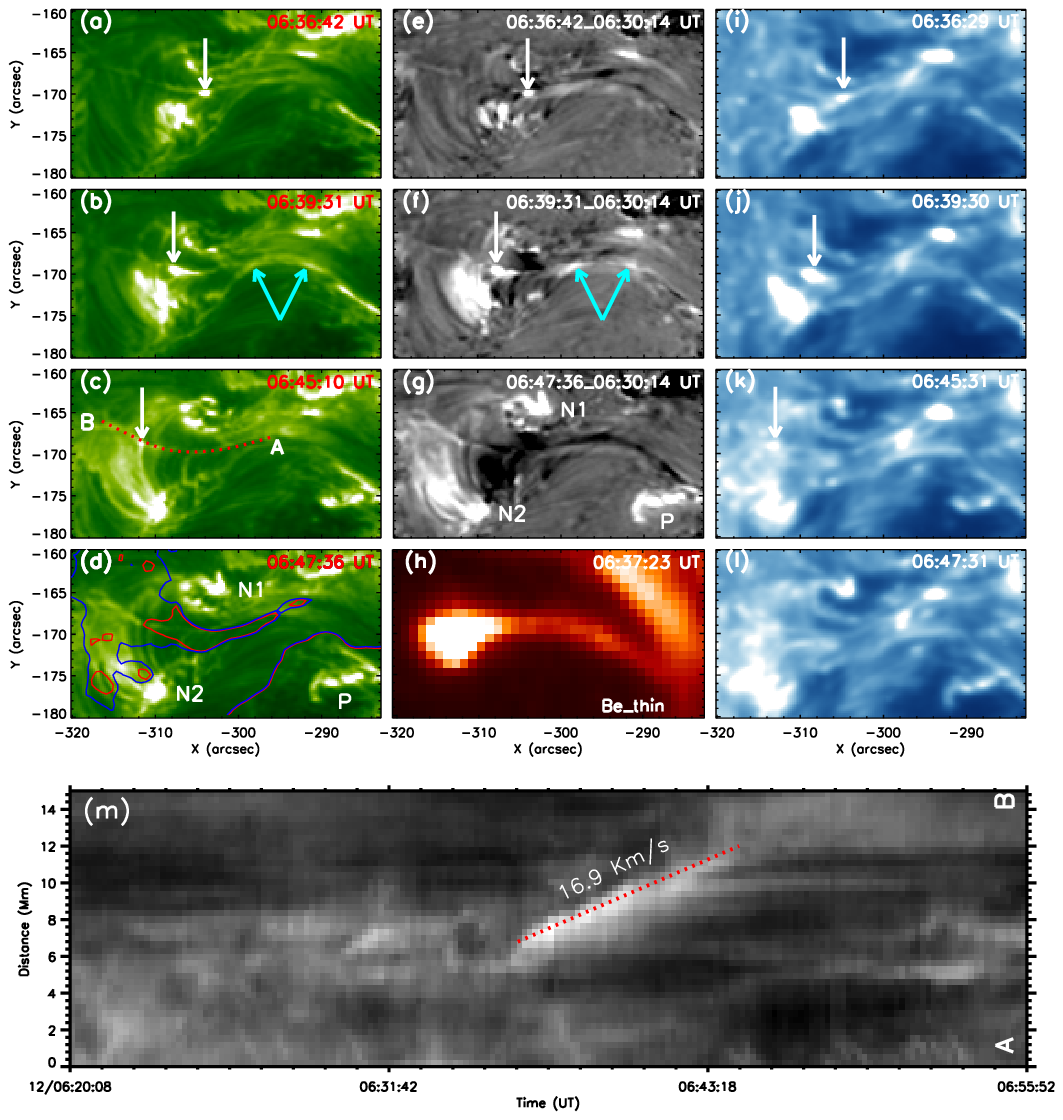}
\caption{Sequence of NVST $H_{\alpha}$  line center (panels ({\it a} \sbond {\it d})), fixed-base difference (panels ({\it e} \sbond {\it g})), and \emph{Hinode}/SOT  \ion{Ca}{2} H (panels ({\it i} \sbond {\it l})) 
 images capturing the genesis and propagation of a compact bright feature along the spine of the fishbone-like structure during a brightening process. 
 The white arrows trace the motion of the compact bright feature, and the cyan arrows indicate mass flows toward the distant positive sunspot P. 
The iso-Gauss contours of  $\pm 50$ G superposed by red and blue curves in panel ({\it d}) are derived from the magnetogram acquired at 06:20:44 UT.
A \emph{Hinode}/XRT Be-thin image (panel ({\it h})) shows the overall morphology of the brightening structure in the solar corona.
Distance-time plots from NVST  $H_{\alpha}$  line center images for the slit ``AB''  is  shown in panel  ({\it m}).
The red dashed line in panel  ({\it m}) outline the motions of the compact brightening. An animation of panels ({\it a}) \sbond ({\it d}) is available. 
The animation has a 16 s cadence, covering 05:34:20 UT to 06:55:52 UT on 2015 March 12.}
\end{figure}

\begin{figure}
\epsscale{1.0}
\plotone{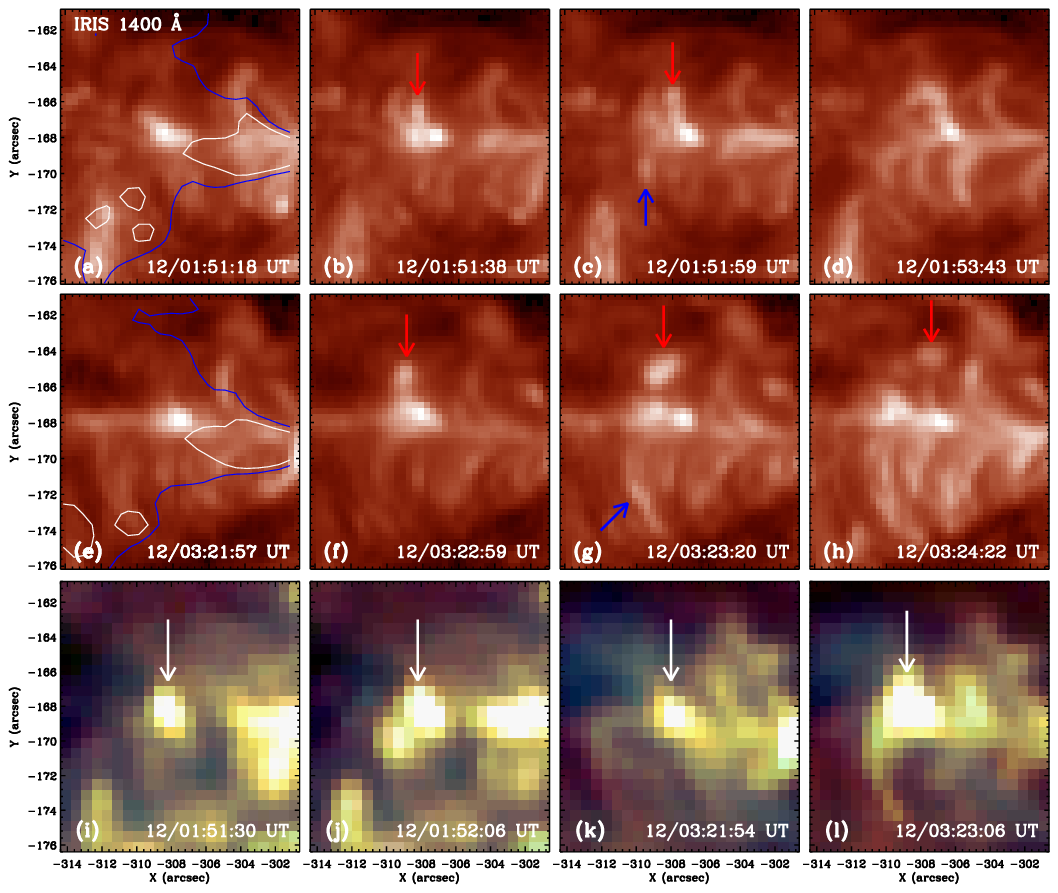}
\caption{\emph{IRIS} 1400 \AA\ (panels ({\it a} \sbond {\it h})) images show mass flows originating from a nodular feature on the spine of the fishbone-like structure. 
The flows move in nearly opposite directions, approximately perpendicular to the spine. 
The red and blue arrows trace the mass flows moving northward and southward, respectively.
Composite of the AIA 171 \AA\ (red), 193 \AA\  (green), and 131 \AA\ (blue) (panels ({\it i} \sbond {\it l})) images  show the coronal brightening processes accompanying the mass flows. 
White arrows denote to the brightenings.
Iso-Gauss contours of  $\pm 100$ G are superposed by white and blue curves in panels ({\it a}) and  ({\it e})).
An animation of panels (({\it a} \sbond {\it d}) and panels (({\it e} \sbond {\it h}) is available. In the movie, events 1\sbond2 correspond to panels (({\it a} \sbond {\it d}) and ({\it e} \sbond {\it h}), respectively.
The animation has a 1 s  cadence.}
\end{figure}

\begin{figure}
\epsscale{1.0}
\plotone{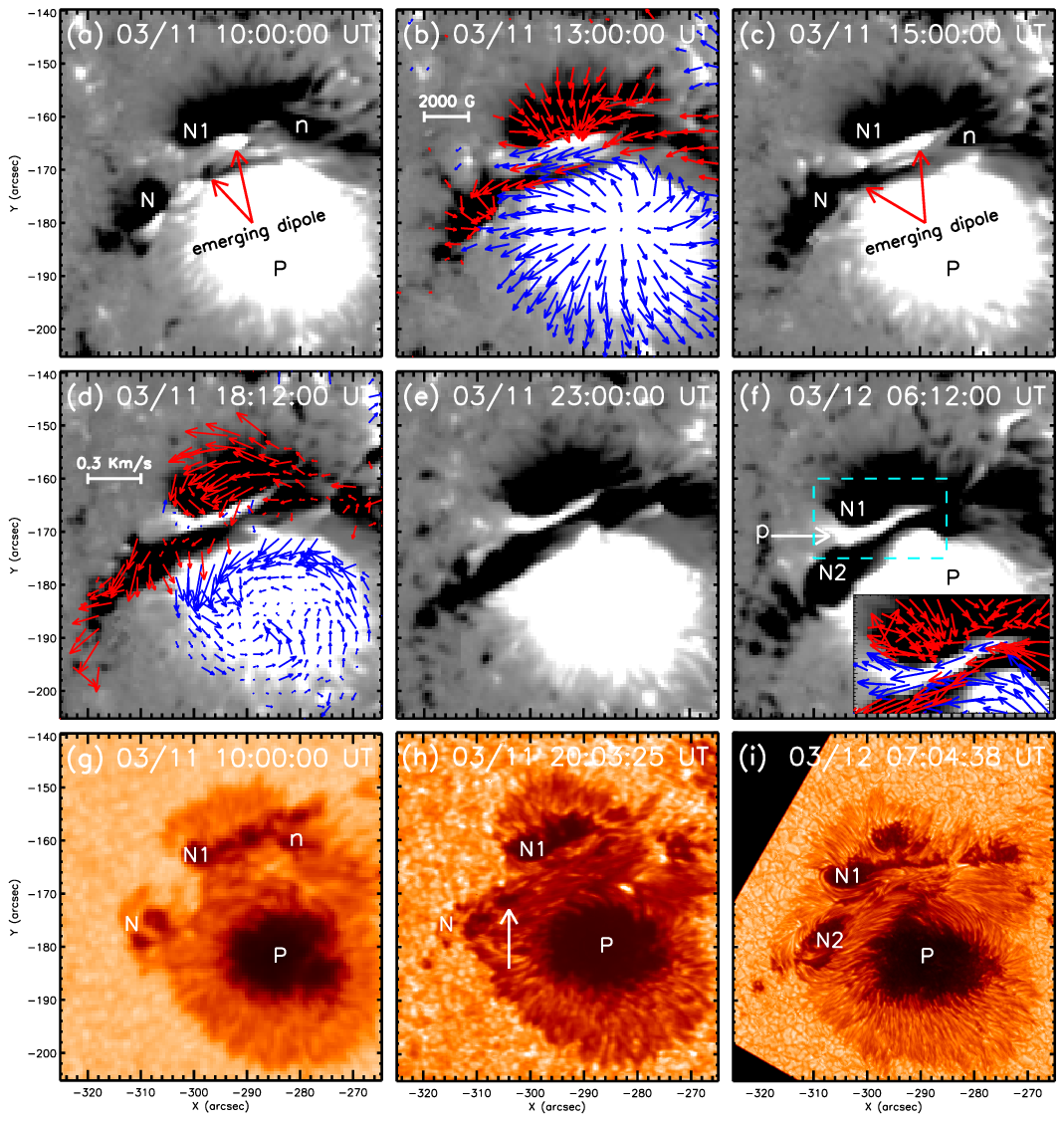}
\caption{Vertical component of the \emph{SDO}/HMI  vector magnetic field (panels ({\it a} \sbond {\it f}) ) showing the evolution of the photospheric magnetic field underlying the fishbone-like structure.
HMI continuum intensity image (panel ({\it g})), \emph{Hinode}/SOT  G-band image (panel ({\it h})), and NVST TiO image (panel ({\it i})) present the evolution of the sunspots. 
The red (blue) arrows in panels ({\it b}) and  ({\it f}) are the horizontal magnetic field vectors and  in panel ({\it d}) are the tangential velocity vectors, which originate from negative (positive) longitudinal fields.
The tangential velocity vectors are inferred from the DAVE4VM technique and averaged from 16:12 UT to 18:12 UT on 2015 March 11. 
In panels  ({\it a}) and  ({\it c}), red arrows trace an emerging  dipole. The cyan dashed box indicates the FOV of the inserted image.
The letters ``P," ``p," ``N1," and ``N2" have the same meaning as in Figure 1({\it a}), ``N" and ``n" denote the sunspots with negative polarity. 
The white arrow in panel  ({\it h}) points to a small pore, corresponding to the negative polarity of the emerging dipole.
An animation of the vertical component of the \emph{SDO}/HMI  vector magnetic field and the continuum intensity image is available.
The animation spans from 00:00 UT on 2015 March 11 to 06:12 UT on 2015 March 12 with a 5 s cadence.}
\end{figure}


\begin{thebibliography}{}

\bibitem[Antolin et al.(2021)]{antolin21} Antolin, P., Pagano, P., Testa, P., et al.\ 2021, Nature Astronomy, 5, 54. doi:10.1038/s41550-020-1199-8
\bibitem[Aschwanden(2019)]{asch19} Aschwanden, M.~J.\ 2019, \apj, 874, 2, 131. doi:10.3847/1538-4357/ab0b42
\bibitem[Aschwanden et al.(2016)]{asch16} Aschwanden, M.~J., Crosby, N.~B., Dimitropoulou, M., et al.\ 2016, \ssr, 198, 1-4, 47. doi:10.1007/s11214-014-0054-6
\bibitem[Aulanier et al.(2006)]{aul06}  Aulanier, G., Pariat, E., D{\'e}moulin, P., et al.\ 2006, \solphys, 238, 2, 347. doi:10.1007/s11207-006-0230-2
\bibitem[Bak et al.(1987)]{bak87} Bak, P., Tang, C., \& Wiesenfeld, K.\ 1987, \prl, 59, 4, 381. doi:10.1103/PhysRevLett.59.381
\bibitem[Bak et al.(1988)]{bak88} Bak, P., Tang, C., \& Wiesenfeld, K.\ 1988, \pra, 38, 364. doi:10.1103/PhysRevA.38.364
\bibitem[Berger \& Asgari-Targhi(2009)]{berger09} Berger, M.~A. \& Asgari-Targhi, M.\ 2009, \apj, 705, 1, 347. doi:10.1088/0004-637X/705/1/347
\bibitem[Bi et al.(2023)]{bi23} Bi, Y., Yang, J.-Y., Qin, Y., et al.\ 2023, \aap, 679, A9. doi:10.1051/0004-6361/202346944
\bibitem[Chae \& Lee(2023)]{chae23} Chae, J. \& Lee, K.-S.\ 2023, \apj, 954, 1, 45. doi:10.3847/1538-4357/ace771
\bibitem[Charbonneau et al.(2001)]{char01} Charbonneau, P., McIntosh, S.~W., Liu, H.-L., et al.\ 2001, \solphys, 203, 2, 321. doi:10.1023/A:1013301521745
\bibitem[Chen et al.(2025)]{chen25} Chen, H., Tian, H., Priest, E.~R., et al.\ 2025, \apj, 995, 1, 94. doi:10.3847/1538-4357/ae12e9
\bibitem[Chen et al.(2020)]{chen20} Chen, H., Zhang, J., De Pontieu, B., et al.\ 2020, \apj, 899, 1, 19. doi:10.3847/1538-4357/ab9cad
\bibitem[Chen(2011)]{chen11} Chen, P.~F.\ 2011, Living Reviews in Solar Physics, 8, 1, 1. doi:10.12942/lrsp-2011-1
\bibitem[Chifor et al.(2008)]{chif08} Chifor, C., Isobe, H., Mason, H.~E., et al.\ 2008, \aap, 491, 1, 279. doi:10.1051/0004-6361:200810265
\bibitem[Chitta et al.(2022)]{chit22} Chitta, L.~P., Peter, H., Parenti, S., et al.\ 2022, \aap, 667, A166. doi:10.1051/0004-6361/202244170
\bibitem[Chitta et al.(2026)]{chit26} Chitta, L.~P., Pontin, D.~I., Priest, E.~R., et al.\ 2026, \aap, 705, A113. doi:10.1051/0004-6361/202557253
\bibitem[Cirtain et al.(2013)]{cir13} Cirtain, J.~W., Golub, L., Winebarger, A.~R., et al.\ 2013, \nat, 493, 7433, 501. doi:10.1038/nature11772
\bibitem[Cheng et al.(2017)]{cheng17} Cheng, X., Guo, Y., \& Ding, M.\ 2017, Science China Earth Sciences, 60, 1383. doi:10.1007/s11430-017-9074-6
\bibitem[Cheng et al.(2023)]{cheng23} Cheng, X., Priest, E.~R., Li, H.~T., et al.\ 2023, Nature Communications, 14, 2107. doi:10.1038/s41467-023-37888-w
\bibitem[De Pontieu et al.(2014)]{de14} De Pontieu, B., Title, A.~M., Lemen, J.~R., et al.\ 2014, \solphys, 289, 7, 2733. doi:10.1007/s11207-014-0485-y
\bibitem[Ding et al.(2022)]{ding22} Ding, T., Zhang, J., \& Hong, J.\ 2022, \apjl, 933, 2, L38. doi:10.3847/2041-8213/ac7c73
\bibitem[D{\'e}moulin et al.(1996)]{demoulin96} D{\'e}moulin, P., Priest, E.~R., \& Lonie, D.~P.\ 1996, \jgr, 101, A4, 7631. doi:10.1029/95JA03588
\bibitem[Dud{\'\i}k et al.(2014)]{dudik14} Dud{\'\i}k, J., Janvier, M., Aulanier, G., et al.\ 2014, \apj, 784, 2, 144. doi:10.1088/0004-637X/784/2/144
\bibitem[Guglielmino et al.(2018)]{gug18} Guglielmino, S.~L., Zuccarello, F., Young, P.~R., et al.\ 2018, \apj, 856, 2, 127. doi:10.3847/1538-4357/aab2a8
\bibitem[Huang et al.(2018)]{huang18} Huang, Z., Xia, L., Nelson, C.~J., et al.\ 2018, \apj, 854, 2, 80. doi:10.3847/1538-4357/aaa9ba
\bibitem[Hood et al.(2016)]{hood16} Hood, A.~W., Cargill, P.~J., Browning, P.~K., et al.\ 2016, \apj, 817, 1, 5. doi:10.3847/0004-637X/817/1/5
\bibitem[Janvier et al.(2013)]{janvier13} Janvier, M., Aulanier, G., Pariat, E., \& D{\'e}moulin, P.\ 2013, \aap, 555, A77. doi:10.1051/0004-6361/201321164
\bibitem[Ji et al.(2019)]{ji19} Ji, K., Liu, H., Jin, Z., Shang, Z., \& Qiang, Z. 2019, ChSBu, 64, 16, 1738. doi:10.1360/N972019-00092
\bibitem[Jiang et al.(2007)]{jiang07} Jiang, Y.~C., Chen, H.~D., Li, K.~J., et al.\ 2007, \aap, 469, 1, 331. doi:10.1051/0004-6361:20053954
\bibitem[Klimchuk(2006)]{klim06} Klimchuk, J.~A.\ 2006, \solphys, 234, 1, 41. doi:10.1007/s11207-006-0055-z
\bibitem[Kobayashi et al.(2014)]{kob14} Kobayashi, K., Cirtain, J., Winebarger, A.~R., et al.\ 2014, \solphys, 289, 11, 4393. doi:10.1007/s11207-014-0544-4
\bibitem[Kosugi et al.(2007)]{kos07} Kosugi, T., Matsuzaki, K., Sakao, T., et al.\ 2007, \solphys, 243, 1, 3. doi:10.1007/s11207-007-9014-6
\bibitem[Lemen et al.(2012)]{lem12} Lemen, J. R., Title, A. M., Akin, D. J., et al. 2012, \solphys, 275, 17
\bibitem[Li et al.(2016)]{li16} Li, L., Zhang, J., Peter, H., et al.\ 2016, Nature Physics, 12, 9, 847. doi:10.1038/nphys3768
\bibitem[Li \& Zhang(2015)]{li15}  Li, T. \& Zhang, J.\ 2015, \apjl, 804, 1, L8. doi:10.1088/2041-8205/804/1/L8
\bibitem[Lin \& Forbes(2000)]{lin2000} Lin, J. \& Forbes, T.~G.\ 2000, \jgr, 105, A2, 2375. doi:10.1029/1999JA900477
\bibitem[Liu et al.(2014)]{liu14} Liu, Z., Xu, J., Gu, B.-Z., et al.\ 2014, Research in Astronomy and Astrophysics, 14, 6, 705-718. doi:10.1088/1674-4527/14/6/009
\bibitem[Lu \& Hamilton(1991)]{lu91} Lu, E.~T. \& Hamilton, R.~J.\ 1991, \apjl, 380, L89. doi:10.1086/186180
\bibitem[M{\"u}ller et al.(2020)]{muller20} M{\"u}ller, D., St. Cyr, O.~C., Zouganelis, I., et al.\ 2020, \aap, 642, A1. doi:10.1051/0004-6361/202038467
\bibitem[Okamoto et al.(2008)]{oka08} Okamoto, T.~J., Tsuneta, S., Lites, B.~W., et al.\ 2008, \apjl, 673, 2, L215. doi:10.1086/528792
\bibitem[Parenti et al.(2010)]{pare10} Parenti, S., Reale, F., \& Reeves, K.~K.\ 2010, \aap, 517, A41. doi:10.1051/0004-6361/200913697
\bibitem[Parker(1957)]{parker57} Parker, E.~N.\ 1957, \jgr, 62, 4, 509. doi:10.1029/JZ062i004p00509
\bibitem[Parker(1983)]{parker83} Parker, E.~N.\ 1983, \apj, 264, 642. doi:10.1086/160637
\bibitem[Parker(1988)]{parker88} Parker, E.~N.\ 1988, \apj, 330, 474. doi:10.1086/166485
\bibitem[Pesnell et al.(2012)]{pes12} Pesnell, W.~D., Thompson, B.~J., \& Chamberlin, P.~C.\ 2012, \solphys, 275, 1-2, 3. doi:10.1007/s11207-011-9841-3
\bibitem[Peter et al.(2022)]{peter22} Peter, H., Chitta, L.~P., Chen, F., et al.\ 2022, \apj, 933, 2, 153. doi:10.3847/1538-4357/ac7219
\bibitem[Polito et al.(2017)]{pol17} Polito, V., Del Zanna, G., Valori, G., et al.\ 2017, \aap, 601, A39. doi:10.1051/0004-6361/201629703
\bibitem[Pontin et al.(2017)]{pontin17} Pontin, D.~I., Janvier, M., Tiwari, S.~K., et al.\ 2017, \apj, 837, 2, 108. doi:10.3847/1538-4357/aa5ff9
\bibitem[Priest \& D{\'e}moulin(1995)]{priest95} Priest, E.~R. \& D{\'e}moulin, P.\ 1995, \jgr, 100, A12, 23443. doi:10.1029/95JA02740
\bibitem[Rast et al.(2021)]{rast21} Rast, M.~P., Bello Gonz{\'a}lez, N., Bellot Rubio, L., et al.\ 2021, \solphys, 296, 4, 70. doi:10.1007/s11207-021-01789-2
\bibitem[Savitzky \& Golay(1964)]{sav64} Savitzky, A., \& Golay, M. J. E. 1964, Anal. Chem., 36, 1627, doi: 10.1021/ac60214a047 
\bibitem[Schou et al.(2012)]{sch12} Schou, J., Scherrer, P.~H., Bush, R.~I., et al.\ 2012, \solphys, 275, 1-2, 229. doi:10.1007/s11207-011-9842-2
\bibitem[Schrijver(2007)]{sch07} Schrijver, C.~J.\ 2007, \apjl, 662, 2, L119. doi:10.1086/519455
\bibitem[Schuck(2008)]{sch08} Schuck, P.~W.\ 2008, \apj, 683, 2, 1134. doi:10.1086/589434
\bibitem[Shen et al.(2024)]{shen24} Shen, Y., Liu, D., Yao, S., et al.\ 2024, \apj, 964, 2, 125. doi:10.3847/1538-4357/ad2349
\bibitem[Shibata \& Magara(2011)]{shibata11} Shibata, K. \& Magara, T.\ 2011, Living Reviews in Solar Physics, 8, 1, 6. doi:10.12942/lrsp-2011-6
\bibitem[Thalmann et al.(2014)]{tha14} Thalmann, J.~K., Tiwari, S.~K., \& Wiegelmann, T.\ 2014, \apj, 780, 1, 102. doi:10.1088/0004-637X/780/1/102
\bibitem[Titov et al.(2002)]{titov02} Titov, V.~S., Hornig, G., \& D{\'e}moulin, P.\ 2002, \jgr, 107, A8, 1164. doi:10.1029/2001JA000278
\bibitem[Tiwari et al.(2014)]{tiw14} Tiwari, S.~K., Alexander, C.~E., Winebarger, A.~R., et al.\ 2014, \apjl, 795, 1, L24. doi:10.1088/2041-8205/795/1/L24
\bibitem[van Ballegooijen \& Martens(1989)]{van89} van Ballegooijen, A.~A. \& Martens, P.~C.~H.\ 1989, \apj, 343, 971. doi:10.1086/167766
\bibitem[Xiang et al.(2016)]{xiang16} Xiang, Y.-. yuan ., Liu, Z., \& Jin, Z.-. yu .\ 2016, \na, 49, 8. doi:10.1016/j.newast.2016.05.002
\bibitem[Xue et al.(2016)]{xue16} Xue, Z., Yan, X., Cheng, X., et al.\ 2016, Nature Communications, 7, 11837. doi:10.1038/ncomms11837
\bibitem[Yan et al.(2017)]{yan17} Yan, X.~L., Jiang, C.~W., Xue, Z.~K., et al.\ 2017, \apj, 845, 1, 18. doi:10.3847/1538-4357/aa7e29
\bibitem[Yan et al.(2022)]{yan22} Yan, X., Xue, Z., Jiang, C., et al.\ 2022, Nature Communications, 13, 640. doi:10.1038/s41467-022-28269-w
\bibitem[Yang et al.(2016)]{yang16} Yang, B., Jiang, Y., Yang, J., et al.\ 2016, \apj, 816, 1, 41. doi:10.3847/0004-637X/816/1/41
\bibitem[Yang et al.(2018)]{yang18} Yang, B., Yang, J., Bi, Y., et al.\ 2018, \apj, 861, 2, 135. doi:10.3847/1538-4357/aac37f
\bibitem[Yang et al.(2021)]{yang21} Yang, B., Yang, J., Bi, Y., et al.\ 2021, \apjl, 921, L33. doi:10.3847/2041-8213/ac31b6
\bibitem[Yang et al.(2023)]{yang23} Yang, L., Yan, X., Xue, Z., et al.\ 2023, \apj, 945, 2, 96. doi:10.3847/1538-4357/acb6f6


\end{thebibliography}
\end{document}